\documentclass[%
twocolumn,
longbibliography,
amsmath,
amssymb,
aps,
pre
]{revtex4-1}

\usepackage{graphicx}
\usepackage{dcolumn}
\usepackage{bm}
\usepackage{siunitx}
\usepackage{tablefootnote}
\usepackage{physics}
\usepackage[dvipsnames]{xcolor}{\huge }
\usepackage[section]{placeins}
\definecolor{darkblue}{rgb}{0,0,.65}
\definecolor{darkgreen}{rgb}{0.28,0.41,0.19}
\usepackage{nicefrac}
\usepackage[%
  pdfauthor={Benedikt Placke},%
  pdfstartview=FitH,%
  breaklinks=true,%
  bookmarks=true,%
  colorlinks=true,%
  anchorcolor=black,%
  citecolor=darkgreen,
  filecolor=black,%
  menucolor=black,%
  urlcolor=darkblue,%
  linkcolor=blue,%
 ]{hyperref}

\def\equationautorefname~#1\null{Eq. (#1)\null}

\newcommand{\appref}[1]{\hyperref[#1]{App.~\ref*{#1}}}
\usepackage{physics}

\begin{document}

\preprint{APS/123-QED}

\title{The Random-Bond Ising Model and its dual in Hyperbolic Spaces}

\author{Benedikt Placke}%
\email{placke@pks.mpg.de}
\affiliation{Max-Planck-Institut f\"ur Physik komplexer Systeme, 01187 Dresden, Germany}
\author{Nikolas P. Breuckmann}%
\email{niko.breuckmann@bristol.ac.uk}
\affiliation{School of Mathematics, University of Bristol, Fry Building Woodland Road BS8 1UG}

\date{\today}

\begin{abstract}
We analyze the thermodynamic properties of the random-bond Ising model (RBIM) on closed hyperbolic surfaces using Monte Carlo and high-temperature series expansion techniques.
We also analyze the \emph{dual}-RBIM, that is the model that in the absence of disorder is related to the RBIM via the Kramers--Wannier duality.
Even on self-dual lattices this model is different from the RBIM, unlike in the euclidean case.
We explain this anomaly by a careful re-derivation of the Kramers--Wannier duality.
For the (dual-)RBIM, we compute the paramagnet-to-ferromagnet phase transition as a function of both temperature $T$ and the fraction of antiferromagnetic bonds~$p$. 
We find that as temperature is decreased in the RBIM, the paramagnet gives way to either a ferromagnet or a spin-glass phase via a second-order transition compatible with mean-field behavior.
In contrast, the dual-RBIM undergoes a strongly first order transition from the paramagnet to the ferromagnet both in the absence of disorder and along the Nishimori line.
We study both transitions for a variety of hyperbolic tessellations and comment on the role of coordination number and curvature.
The extent of the ferromagnetic phase in the dual-RBIM corresponds to the correctable phase of  hyperbolic surface codes under independent bit- and phase-flip noise. 
\end{abstract}

\maketitle

\section{Introduction}

The effect of quenched disorder to critical phenomena in spin systems has been the subject of intense study for almost half a century. One of the central models has been the random bond Ising model (RBIM), which serves as a model for certain spin glass materials \cite{Edwards1975, binder1986}, certain localization problems and plateau transitions in the quantum Hall effect \cite{cho1997,merz2002} but also has been shown to be relevant for the analysis of the performance of topological quantum error correcting codes when assuming certain noise models \cite{dennis2002topological, chubb2021statistical,wang2003confinement,kubica2018three,kovalev2014spin,jiang2019duality}. 

The RBIM in flat space has been understood quite comprehensively by now: while weak disorder is irrelevant in the renormalization group sense \cite{dotsenko1983}, increasing the disorder strength lowers the phase transition temperature until the so called ``Nishimori point'' is reached. Beyond this, the system stays disordered for all temperatures. In more than two dimensions, the system for low temperatures and large disorder is in a spin glass phase, with the Nishimori point being the tri-critical point.

The present paper is now concerned with the properties of the RBIM in curved space.
Condensed matter physics in curved spaces has been a subject of intense study. 
Curvature is known, for example, to alter the critical properties of statistical mechanics models \cite{callan1990infrared}, circuit quantum electrodynamics \cite{Kollar2019hyperbolic,bienias2022circuit} and band theory \cite{maciejko2021hyperbolic, boettcher2022crystallography, attar2022selberg, Ikeda2021hyperbolic, Lenggenhager2022simulating}.
The Ising model in curved space has, to the best of our knowledge, so far only been studied in the absence of disorder \cite{Rietman1992, krcmar2008, mnasri2015, jiang2019duality, breuckmann2020}. 
In this limit, the  model undergoes a phase transition from a paramagnetic high-temperature to a low-temperature ferromagnetic phase, just as its flat-space counterpart. 
The transition is mean-field in nature, but surprisingly it is not located at the fixed-point of the Kramers--Wannier duality, even on self-dual hyperbolic lattices. This observation implies either the existence of a second phase transition, for which no evidence was found numerically, or a violation of self-duality of the Ising model on self-dual hyperbolic lattices. 
We note that the existence of a second phase transition for the pure Ising model on the hyperbolic plane with free boundary condition has been proved~\cite{Wu1996, Wu2000,jiang2019duality}.

Studying the Kramers--Wannier duality in the presence of curvature is interesting on its own right \cite{freed2018topological}. However, as Polyakov pointed out already in 1987 \cite[Chapter 9]{polyakov1987gauge}, its understanding will have consequences also for related constructions. This includes Polyakov's original example, the Fermionization of Ising spins \cite{dotsenko1988fermion} and, more recently, the mapping between the decoding of homological quantum error correction codes and statistical mechanics models \cite{dennis2002topological, chubb2021statistical,wang2003confinement,kubica2018three,kovalev2014spin,jiang2019duality}.

As we show in this work, there is an anomaly in the hyperbolic RBIM.
It turns out that it is not self-dual even on self-dual lattices, but, in the disorder-free limit, is related by the Kramers--Wannier duality to what we call the \emph{dual-RBIM}.
Hence, in this paper we study both the critical properties of the random bond Ising model and its dual in hyperbolic space.
Note that what we call the dual-RBIM it is not related to the RBIM by an exact duality in the presence of disorder.

We begin our study of both models by mapping out their phase diagrams using a combination of high-temperature series expansion techniques and Monte-Carlo simulations. We show that the RBIM realizes a paramagnetic, a ferromagnetic and a spin-glass phase with the Nishimori point as the tricritical point. All transitions (with the exception of the multicritical point) are compatible with second-order mean-field behavior. In contrast, the dual-RBIM in the disorder-free limit as well as along the Nishiori line undergoes a strongly first-order transition as evidenced through Metropolis and canonical simulations using the Wang-Landau algorithm.
We numerically verify the duality of the two models in the disorder-free case and show that a duality conjectured by Takeda et al. \cite{takeda2005exact} is fulfilled only approximately.

The rest of the paper is organized as follows. In \autoref{sec:intro_rbim} we give necessary notions and definitions; in particular the dual-RBIM is derived in \autoref{sec:intro_duality}. In \autoref{sec:methods} we derive the high-temperature expansion for the RBIM and give details on the Monte-Carlo simulations used. \autoref{sec:rbim} presents the results on the phase diagram and critical properties of the random Bond Ising model and \autoref{sec:dual-rbim} presents the same for the dual model. Finally, we discuss the relevance of our results to the decoding of hyperbolic surface codes in \autoref{sec:qec}. We conclude in \autoref{sec:conclusion}.

\section{The disordered Ising model and its dual in the hyperbolic plane\label{sec:intro_rbim}}

\subsection{Hyperbolic surfaces}\label{sec:intro_hyperbolic}

\begin{figure}
    \centering
    \includegraphics[width=0.65\columnwidth]{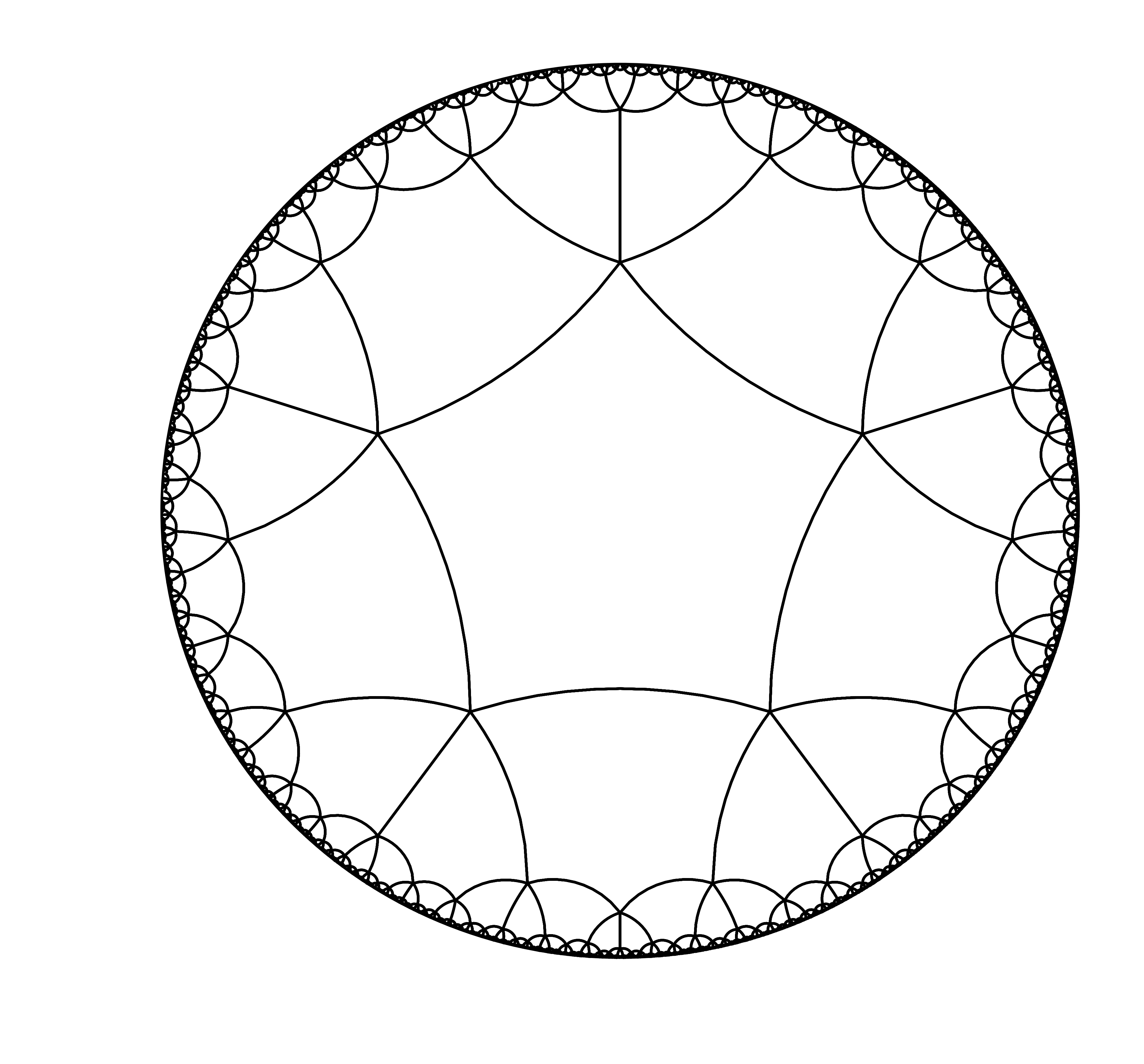}
    \caption{(a) Poincar\'e disk model of the infinite hyperbolic plane $\mathbb{H}^2$ with the $\{5,5\}$ lattice. All edges have the same length with respect to the hyperbolic metric, see \autoref{eq:metric}.}
    \label{fig:55tessellation}
\end{figure}

The hyperbolic plane is a 2D manifold of constant negative curvature.
It can be realized in terms of several models.
Here, we will employ the \emph{Poincar\'e disk model}, which is defined as follows.
Consider a disk in $\mathbb{R}^2$ with unit radius and centered at the origin.
Let $x$ and $y$ denote the standard coordinates of $\mathbb{R}^2$.
Then the hyperbolic plane is given by the set of points
\begin{align}
    \mathbb{H}^2 = \{ (x,y)\in \mathbb{R}^2 \mid x^2+y^2 < 1 \}
\end{align}
with metric given by
\begin{align}\label{eq:metric}
    ds^2 = \frac{dx^2 + dy^2}{\left( 1-x^2-y^2 \right)^2}
\end{align}
It is immediate from \autoref{eq:metric} that length scales are highly distorted towards the boundary of the disk compared to the euclidean metric, see \autoref{fig:55tessellation}.

Just as regular euclidean space can be tessellated by squares, triangles or hexagons, hyperbolic space can be tessellated by regular polygons as well.
In fact, it turns out that hyperbolic space supports an infinite number of regular tessellations.
We can label regular tessellations by the \emph{Schl\"afli symbol} $\{r,s\}$, where $r$ is the number of sides of the polygonal plaquettes and $s$ is the number of plaquettes meeting at each vertex.
For example, the hexagonal lattice has Schl\"afli symbol $\{6,3\}$.
Its dual lattice can be obtained by reversing the Schl\"afli symbol, i.e. the triangular lattice $\{3,6\}$.
These two examples, together with the self-dual square tessellation $\{4,4\}$ are all the possible regular tessellations of the euclidean plane.
The hyperbolic plane supports any regular tessellation $\{r,s\}$ as long as $1/r+1/s < 1/2$.
The $\{5,5\}$ tessellation of the hyperbolic plane in the Poincar\'e disk model is shown in \autoref{fig:55tessellation}.

In order to approximate the infinite hyperbolic plane for numerical analysis, we can consider sequences of finite neighborhoods $B_R$ (discs) of increasing radii $R$.
This is commonly done in the context of statistical mechanics models in euclidean space for performing finite size analysis.
The models differ at the boundaries of the finite regions from the infinite euclidean plane.
However, the effects of this deviation vanish in the thermodynamic limit as $\operatorname{vol}(\partial B_R) / \operatorname{vol}(B_R) \rightarrow 0$ for $R\rightarrow \infty$.
This is not the case in hyperbolic space where $\operatorname{vol}(\partial B_R)$ and $\operatorname{vol}(B_R)$ have the same asymptotic scaling.
This means that taking finite neighborhoods with boundaries can not be used to analyze the behaviour of the infinite model.
We solve this problem by considering families of boundaryless, finite surfaces (supporting the same tessellation) which are indistinguishable from the infinite hyperbolic plane in local regions of increasing size at any point.

\begin{figure}[h]
    \centering
    \includegraphics[width=0.2\textwidth]{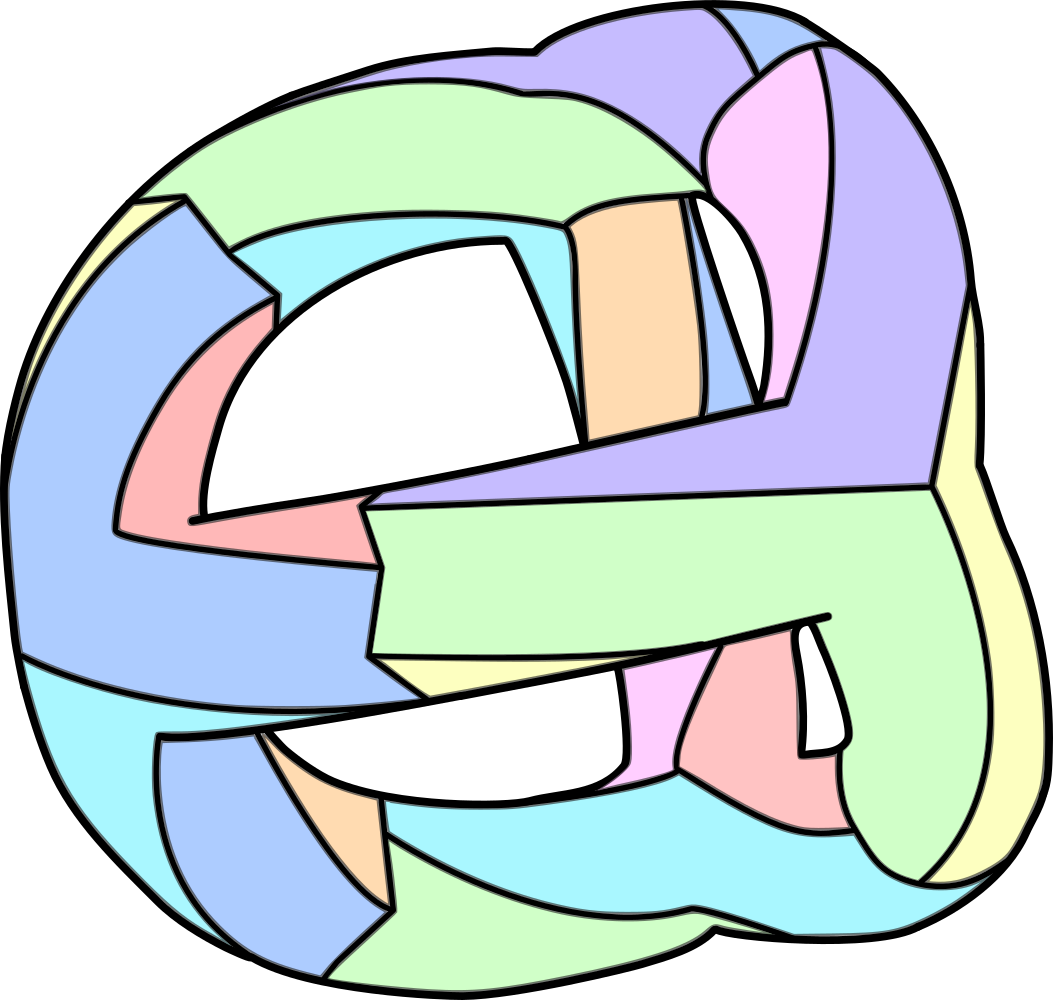}
    \hfil
    \includegraphics[width=0.2\textwidth]{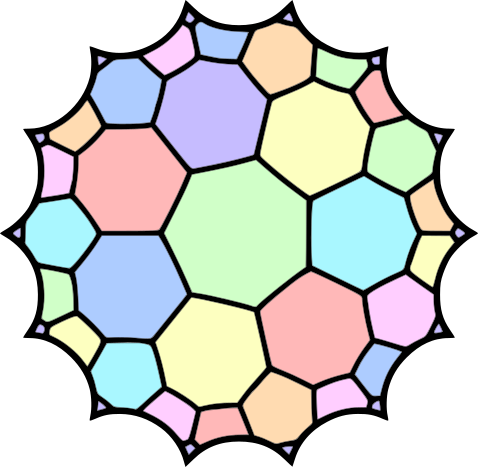}
    \caption{A hyperbolic surface of genus 3 tessellated by the $\{7,3\}$-tessellation (left).
    If we cut the surface open we obtain a flat piece of hyperbolic space (right).
    The plaquettes are colored to guide they eye.}
    \label{fig:klein_quartic}
\end{figure}

Introducing periodic boundary conditions is a much more subtle process in hyperbolic spaces compared to euclidean spaces. 
In particular, closed, orientable hyperbolic manifolds have a genus that is proportional to their area.
This is seen most easily by considering a theorem due to Gau\ss --Bonnet, which states that the geometry (curvature) of a 2D surface is connected to its topology.
More concretely, it states that for any orientable surface~$S$ of genus~$g$ it holds that
\begin{align}\label{eqn:gauss_bonnet}
    2-2g = \frac{1}{2\pi} \int_S \kappa\, dA 
\end{align}
where on the right hand side we integrate the curvature~$\kappa$ at every point in $S$ over the area of $S$.
If $S$ is euclidean, then the curvature~$\kappa$ is equal to 0 at every point.
From \autoref{eqn:gauss_bonnet} it then immediately follows that all orientable euclidean surfaces are tori ($g=1$).
On the other hand, if $S$ is hyperbolic then $\kappa = -1$ everywhere. Orientable hyperbolic surfaces hence have
\begin{align}\label{eq:area_propto_genus}
    \frac{\operatorname{area}(S)}{2\pi} = 2g-2
\end{align}
so that larger surfaces necessarily have a higher genus.
In \autoref{fig:klein_quartic} we show an example of a closed $g=3$ hyperbolic surface, called \emph{Klein quartic}, which supports a $\{7,3\}$ tessellation.

As it turns out, the subtlety that hyperbolic surfaces are topologically complex becomes important in the Kramers--Wannier duality.
This is because the Kramers--Wannier duality is sensitive to the number of closed loops (cycles) in the lattice and the higher genus of hyperbolic surfaces introduces more such loops, see discussion in \autoref{sec:intro_duality}.

\subsection{Duality in the Hyperbolic Ising model\label{sec:intro_duality}}
We consider the Ising model (for the time being \emph{without} quenched disorder) on a lattice $\mathcal L = (V, E, F)$. We denote by $V$ the set of vertices, by $E$ the set of edges and by $F$ the set of faces of the lattice.
Denoting nearest neighbor bonds between two vertices $i$ and $j$ of the lattice by $\expval{ij}$, the Hamiltonian of the Ising model is then given by
\begin{equation}
\label{eq:pure-Ising}
    H = J \sum_{\expval{ij}} \sigma_i \sigma_j,
\end{equation}
where $\sigma\in\{\pm1\}$ are Ising spin variables and we asume $J<0$ for ferromagnetic coupling.

In euclidean space, the Kramers--Wannier duality \cite{kramers1941} relates the high-temperature expansion of the Ising model [\autoref{eq:pure-Ising}] to its low-temperature expansion of the same model on the dual lattice. 

In particular, Kramers and Wannier showed a exact relation the two partition functions
\begin{subequations}
\label{eq:krammers-wannier}
\begin{align}
    Z(T) = \Tilde Z(T^*)
\end{align}
where~$Z$ and~$\Tilde Z$ are the partition functions of the Ising model on the lattice and its dual respectively, and $T$ and~$T^*$ satisfy 
\begin{equation}
\label{eq:krammers-wannier-T}
    \sinh(2J/T)\sinh(2J/T^*) = 1.
\end{equation}
\end{subequations}

On a self-dual lattice, $Z=\Tilde Z$ and thus the duality [\autoref{eq:krammers-wannier}] constitutes an exact mapping between the behavior of the system at high and low temperature. In particular, assuming that a single phase transition occurs, this fixes the critical temperature to the fixed-point of \autoref{eq:krammers-wannier-T}
\begin{equation}
    \label{eq:krammers-wannier-Tc}
    \sinh(2J/T_c)\sinh(2J/T_c) = 1~\Rightarrow~T_c \approx 2.2692J.
\end{equation}

An open question posed by earlier studies \cite{Rietman1992, breuckmann2020} was how \autoref{eq:krammers-wannier-Tc} is violated in hyperbolic space. 
That is, if the Kramers--Wannier duality [\autoref{eq:krammers-wannier}] applies also to hyperbolic lattices, one of the following must hold: either all self-dual hyperbolic lattices (that is tesselations of compact hyperbolic manifolds with Schl\"afli symbol $\{r, s\}$ with $r=s$) have the \emph{same} critical temperature, given by \autoref{eq:krammers-wannier-Tc}, or there exist \emph{two} phase transitions, related by \autoref{eq:krammers-wannier-T}.
In fact, as we will show below, the Ising model on tessellations of compact hyperbolic manifolds is not related by the Krammers--Wannier duality to the same Ising model on the dual lattice. In particular, it is not self-dual, even on self-dual lattices.

\subsubsection{Re-derivation of the Kramers--Wannier duality}
To understand this, let us perform a careful re-derivation of the Kramers--Wannier duality.
To this end, we first consider the high-temperature expansion of the Ising model on a lattice $\mathcal L = (V, E, F)$.
Let $Z_1$ be the set of subsets $\gamma \subset E$ such that in the subgraph induced by any such $\gamma$, every vertex has even degree.
The subsets $\gamma \in Z_1$ are called \emph{cycles}.
It is well-known that the partition function can be written as a sum over the set of all cycles of the graph (see e.g.  \cite[Chapter~2]{oitmaa2006}):
\begin{subequations}
\label{eq:kw-high-T}
\begin{align}
		Z(K) &= \sum_{{\sigma}\in \{\pm 1\}^N} \prod_{(i,j)\in E} \exp(K \sigma_i \sigma_j)\\
		&= (\cosh K)^{|E|} \sum_{{\sigma}} \prod_{(i,j)} (1+ \sigma_i \sigma_j \tanh K)\\
		&= 2^N (\cosh K)^{|E|} \sum_{\gamma \in Z_1} (\tanh K)^{|\gamma|}
  \end{align}
\end{subequations}
where we have defined $K=-J/T$ and $|S|$ denotes the size of the set $S$.

\begin{figure}[h]
    \centering
    \includegraphics[width=0.2\textwidth]{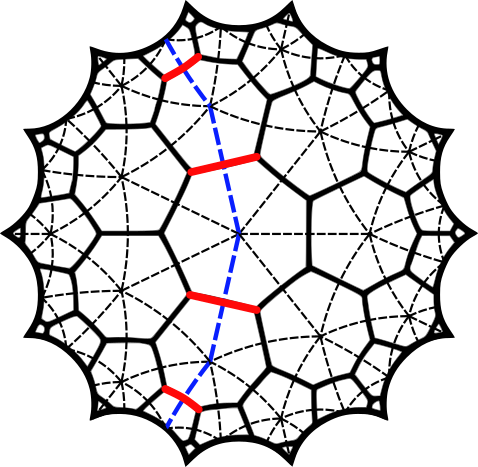}
    \hfil
    \includegraphics[width=0.2\textwidth]{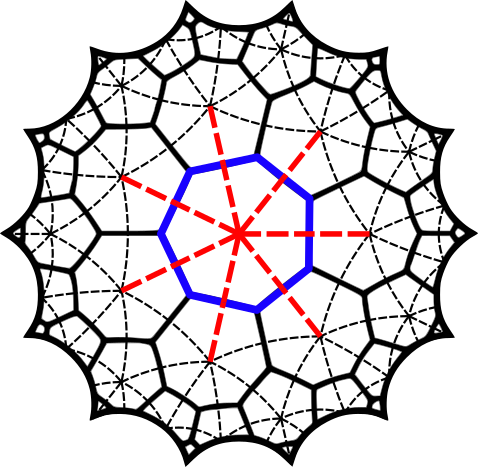}
    \caption{The left shows a cycle on the dual lattice (blue) and the associate cocylce on the primal lattice (red). The right shows a boundary on the primal lattice (blue) and the associate coboundary on the dual lattice (red).}
    \label{fig:klein_quartic_dual}
\end{figure}

Note that the set $Z_1$ of cycles $\gamma$ in \autoref{eq:kw-high-T} includes ones that are contractible as well as ones that are non-contractible. Two examples for such cycles, on a surface with genus 3, tessellated by the $\{7, 3\}$ tessellation (cf. \autoref{fig:klein_quartic}), are given in \autoref{fig:klein_quartic_dual}. On the right we show a contractible cycle on the primal lattice (solid lines) in blue. On the left we show, also in blue, a non-contractible cycle on the dual lattice (dashed lines).

To establish the duality, we also consider the low-temperature expansion of the Ising model, but on the dual lattice $\mathcal L^* = (V^*,E^*,F^*) = (F,E,V)$. For regular tessellations of hyperbolic surfaces, the dual lattice is just obtained by swapping the first and second entry of its Schl\"afli symbol $\{r, s\}$. This is also indicated in \autoref{fig:klein_quartic_dual}. The primal lattice (solid lines) is the $\{7, 3\}$ tessellation and its dual (dashed lines) is the $\{3, 7\}$ tessellation of the same surface.

The low temperature expansion follows from expressing the partition function in terms of excitations on top of the (ferromagnetic) ground state. These are given by domain walls.
For example, consider starting from a all-ferromagnetic state of the Ising model [\autoref{eq:pure-Ising}] on the (dual) lattice indicated by dashed lines in \autoref{fig:klein_quartic_dual}. The cost of flipping the spin on the central cite is given by the size of the domain wall indicated in red on the right of \autoref{fig:klein_quartic_dual}.
Generally, let $B^{1*}$ be the set of all possible domain walls on the dual lattice. We can write
\begin{subequations}
\label{eq:low-T-z}
\begin{align}
\Tilde Z(K)    &= \sum_{{\sigma}\in \{\pm 1\}^N} \prod_{(i,j)\in E^*} \exp(K \sigma_i \sigma_j)\\
&= 2 \sum_{\omega^* \in B^{1*}} \exp(K^*)^{|E^*|-2|\omega^*|}\\
&= 2 \exp(K)^{|E^*|} \sum_{\omega^* \in B^{1*}} \exp(-2 K)^{|\omega^*|}
\end{align}
\end{subequations}
where the second equality directly follows from the definition of $B^{1*}$. In the language of homology, the set $B^{1*}$ is given exactly by the set of \emph{coboundaries} on the \emph{dual} lattice.

The basis of the Kramers--Wannier duality, homologically speaking, is the fact that the set of cycles $Z_1$ is in one-to-one correspondence with the set of \emph{cocycles} $Z^{1*}$ on the dual lattice $\mathcal L^*$. This is also indicated in \autoref{fig:klein_quartic_dual} where we show two examples of the correspondence of cocycles (red) and cycles (blue). The left side shows a non-contractible cocycle on the primal lattice (solid, red) and the corresponding cycle on the dual (blue, dashed). The right side shows a contractible cycle (a \emph{boundary}) on the primal lattice (red, solid) and the corresponding cocycle (a \emph{coboundary}) on the dual lattice (red, dashed). 

Using this equivalence, $Z_1 = Z^{1*}$, as well as \autoref{eq:krammers-wannier-T}, and defining $K^*=J/T^*$, we can then rewrite 
\begin{align}
    Z(K) &= 2 \exp(K^*)^{|E^*|}\sum_{\gamma^* \in Z^{1*}}\exp(-2K^*)^{|\gamma^*|}.
    \label{eq:dual-z}
\end{align}
Above, the right hand side is \emph{almost} the low-temperature expansion of the Ising model on the dual lattice [\autoref{eq:low-T-z}], at temperature $T^*$ [\autoref{eq:krammers-wannier-T}].
The difference between \autoref{eq:dual-z} and \autoref{eq:low-T-z} is that the sum above is over all cocycles $\gamma^* \in Z^{1*}$ whereas the low-temperature expansion is a sum over domain walls $\omega^* \in B^{1*}$, that is coboundaries or ``contractible'' cocycles.
Physically, we can rationalize this difference by looking at the example of a non-contractible cocycle on the left of \autoref{fig:klein_quartic_dual} (red, solid). The corresponding cycle (blue, dashed) appears in the high-temperature expansion of the dual lattice (every vertex in it has even degree). However, there is no set of spins on vertices of the primal lattice that we could flip to get a domain of that form.

Hence, for Ising models on regular tessellations of closed manifolds, we have established what is the \emph{difference} between their high-temperature expansion [\autoref{eq:kw-high-T}] and the low-temperature expansion on the dual lattice, at the dual temperature [\autoref{eq:dual-z}]. In the following we will show that (i) for tessellations of closed euclidean surfaces (tori), this difference vanishes in the thermodynamic limit, yielding the Kramers--Wannier duality [\autoref{eq:krammers-wannier}], and (ii) the difference does \emph{not} vanish for tessellations of closed hyperbolic surfaces, leading to a violation of \autoref{eq:krammers-wannier}.

Note that the contribution of any cocycle in \autoref{eq:dual-z} has a weight $\exp(-2K^*)^{|\gamma^*|}$.
For euclidean lattices on an $L\times L$ torus this implies that the contribution of any non-contractible cocycle is at least of order $\order{\exp(-2K^*)^L}$.
Focussing on such minimal-size cocycles, of which there are $\sim L$, the difference between \autoref{eq:dual-z} and the low-temperature expansion of the Ising model vanishes in the thermodynamic limit
\begin{equation}
    Z(T) - \Tilde Z(T^*) \sim L \exp(-2K^*L) \xrightarrow[L\to \infty]{} 0.
\end{equation}
This then yields \autoref{eq:krammers-wannier}.

In contrast, in hyperbolic space, the number of minimal, non-contractible cocycles goes as $\sim N$ [see \autoref{eq:area_propto_genus}] while their length grows only logarithmically \cite{macaj2008injectivity,moran1997growth}. This means that the same difference goes as
\begin{equation}
    Z(T) - \Tilde Z(T^*) \sim N^{1-2K^*}
\end{equation}
which does not generally vanish as $N\to\infty$. 

\subsubsection{The dual Ising model in hyperbolic space}

In order to obtain a model that does fulfill the Kramers--Wannier duality, we have to define a model where possible domain walls on top of the ferromagnetic ground state include all non-contractible cocycles.

We achieve this by a rather simple trick. Given an Ising model [\autoref{eq:pure-Ising}] on a tessellation of a closed hyperbolic surface~$S$ with $2g$ nonequivalent, non-contractible cocyles~$\ell$, we introduce one additional Ising degree of freedom~$\eta_\ell$ per nonequivalent, non-contractible cocycle.

We then define the ``dual Ising model'' as
\begin{align}\label{eq:H-dual-ising-pure}
    H &= J \sum_{\expval{ij}} \left(  \prod_{\ell\,|\,\expval{ij} \in \ell} \eta_\ell \right)\,\sigma_i \sigma_j
\end{align}
where $J<0 $ as before is chosen to be ferromagnetic  and we have chosen one representative per nontrivial cocycle~$\ell$. One example of such a representative on a hyperbolic surface with genus 3, tessellated by the $\{7, 3\}$-tesselation, is shown on the left side of \autoref{fig:klein_quartic_dual}, where it is highlighted in red. 
The effect of flipping this Ising degree of freedom $\eta_\ell \to -\eta_\ell$ is to reverse the sign of the coupling of each edge that is part of the representative $\ell$. One can think of each variable $\eta_\ell$ to encode the \emph{boundary} condition in one possible direction which can either be periodic ($\eta_\ell=1$) or anti-periodic ($\eta_\ell=-1$). 
Because of this, the domain walls of the model defined by \autoref{eq:H-dual-ising-pure} include the nontrivial cocycles of the lattice and its partition is given by \autoref{eq:dual-z}, that is the dual-RBIM for $p=0$ is indeed the Kramers--Wannier dual of the Ising model.

This model also gives another rational for the difference between duality of Ising models on tessellations of eucledian and hyperbolic manifolds. Strictly speaking, the Kramers--Wannier dual of the Ising model on \emph{finite} tessellations of euclidean manifolds is also given by \autoref{eq:H-dual-ising-pure}. However since all closed, orentable euclidean manifolds are tori, the dual model has only two additional degrees of freedom $\ell$ compared to the original Ising model [\autoref{eq:pure-Ising}]. Hence, they have no finite entropic contribution in the thermodynamic limit and the dual model has the same thermodynamic properties as the original model. In contrast, on tessellations of closed hyperbolic manifolds, the number of additional variables $\ell$ in \autoref{eq:H-dual-ising-pure} is extensive ($\sim N$) and hence changes the properties of the model, even in the thermodynamic limit.

\subsection{The Random-Bond Ising model}

The random-bond Ising model (RBIM), first introduced by Edwards and Anderson \cite{Edwards1975} to model the interaction of dilute magnetic alloys, serves as a simple model to study critical phenomena in systems with quenched disorder.
The Hamiltonian for the RBIM on a lattice with nearest-neighbor bonds $\expval{ij}$ is
\begin{align}
    H = \sum_{\langle i,j \rangle} J_{ij} \sigma_i \sigma_j
    \label{eq:H-RBIM}
\end{align}
where $\sigma_i\in \lbrace \pm 1 \rbrace$ are Ising spin variables and $J_{ij}$ are random couplings.
Whenever we refer to the Ising model in ``hyperbolic space'' or on ``hyperbolic lattices'' throughout this work, we refer to a model where spins are located on the vertices of regular tessellations of \emph{compact} hyperbolic manifolds, with Schl\"afli symbol $\{r,s\}$. This emphasis is important, since considering the same model on non-compact hyperbolic manifolds with, for example, open or closed boundary conditions will generally change its properties \cite{Wu1996, Wu2000}. 
The couplings are distributed independently and identically. In this paper, we take their individual probability distribution to be the so called 
``$\pm J$-distribution''
\begin{align}\label{eq:disorder}
    P(J_{ij}) = p\, \delta(J_{ij}-1) + (1-p) \, \delta(J_{ij}+1)
\end{align}
so that each coupling is anti-ferromagnetic $J_{ij} = +1$ with probability $p$ and ferromagnetic $J_{ij} = -1$ with probability $1-p$.
Hence, on the infinite hyperbolic plane, $p$ is equal to the fraction of anti-ferromagnetic bonds.  
The free energy of the model, when considering quenched disorder is then given by
\begin{align}
    F &= \left[\log(Z) \right], \\
    Z &= \sum_{\{\sigma\}} \exp\left(-\beta \sum_{\langle i,j \rangle} J_{ij} \sigma_i \sigma_j \right),
\end{align}
where brackets $[\dots]$ denote the average over disorder configurations.

For $p=0$, the model reduces to the ferromagnetic Ising model, which we have studied for regular tessellations of compact hyperbolic manifolds in a previous paper \cite{breuckmann2020}. This model as a function of temperature undergoes a phase transition from a high-temperature paramagnetic into a low-temperature ferromagnetic phase. Our study revealed that this transition is mean-field in nature for all investigated tessellations. In the present work, we extend our previous work to the case of finite $0 < p < 1/2$. 

We also study the \emph{dual} Ising model \autoref{eq:H-dual-ising-pure} in the presence of quenched disorder. In this case it becomes
\begin{align}\label{eq:H-dual-ising}
    H &= \sum_{\expval{ij}} J_{ij} \left(  \prod_{\ell\,|\,\expval{ij} \in \ell} \eta_\ell \right)\,\sigma_i \sigma_j.
\end{align}
As before, the $\sigma_j\in\{\pm1\}$ are Ising variables, as are the $\eta_\ell \in \{\pm 1\}$. While the $\sigma_j$ are located on the vertices of the lattice, each $\eta_\ell$ is associated with a nontrivial cocycle $\eta_\ell$ (cf. \autoref{sec:intro_duality}). The $J_{ij}$ are random couplings drawn from the $\pm J$ distribution defined in \autoref{eq:disorder}.

The Kramers--Wannier duality [\autoref{eq:krammers-wannier}], as usual, is only valid is the disorder-free case.
However there is a conjecture by Takeda and Nihsimori \cite{takeda2005exact}  relating the location of the Nishimori point of the RBIM with the position in the dual model
\begin{equation}\label{eq:duality-conjecture}
    H(p_{\rm N}) + H(p_{\rm N}^*) = 1
\end{equation}
where $H(p) = -p \log_2(p) - (1-p)\log_2(1-p)$ is the binary entropy.
As discussed in \autoref{sec:dual-rbim} we see that the conjecture holds approximately, but not within error bars.

\subsection{Possible Phases and Order Parameters}

\begin{figure}
    \centering
    \includegraphics{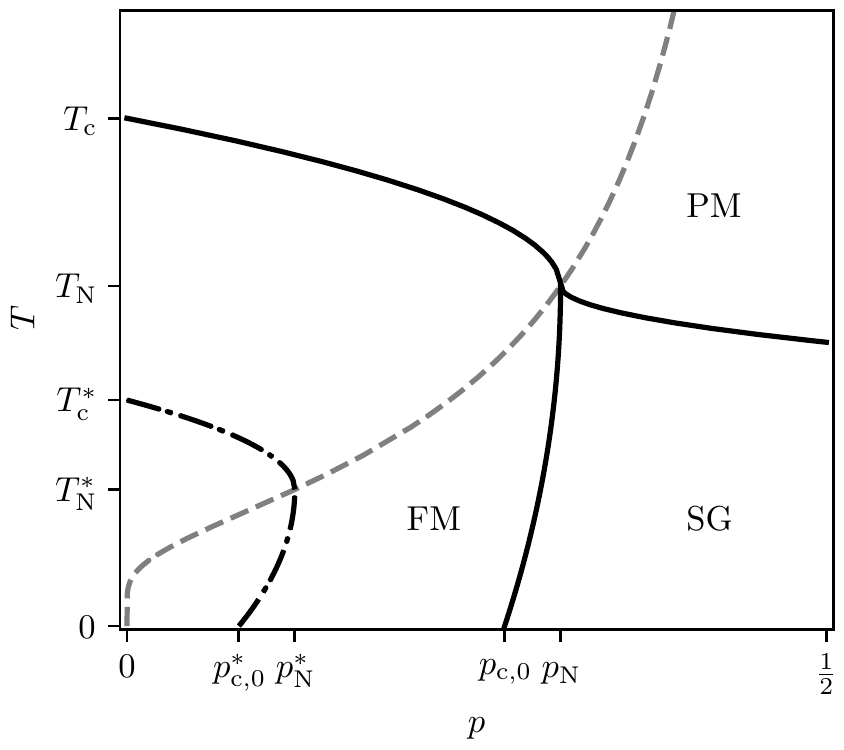}
    \caption{Schematic phase diagram of the random bond Ising model and its dual on the hyperbolic plane as a function of temperature $T$ and the fraction of antiferromagnetic bonds~$p$. The high-temperature paramagnetic (PM) phase at low temperatures gives way either to a ferromagnetic (FM) phase spin glass (SG) phase at weak and strong disorder respectively. 
    For the dual model, we only indicated the schematic boundary of the FM phase.
    Note that although the temperatures~$T_c$ and~$T_c^*$ are related by the Kramers--Wannier relation, the dual model of the hyperbolic Ising model is different from the original model even on self-dual lattices (see main text for details). 
    The phase boundary of the dual model corresponds to the decoding threshold of the hyperbolic surface code under phenomenological noise. The Nishimori line is indicated in dashed-gray. }
    \label{fig:phases_sktech}
\end{figure}

At high temperature, both the RBIM and its dual are in the paramagnetic phase. As the temperature is lowered, at low disorder this gives way to a ferromagnetic phase which is continuously connected to that of the pure model at $p=0$. The transition from the paramagnet to the ferromagnet corresponds to an instability of the mean of the magnetization distribution $\rho(m)$. That means while in the paramagnet we have 
\begin{equation}
    \rho(m) = \delta(m),
\end{equation}
in the ferromagnetic phase 
\begin{equation}
    \rho(m) = \delta(\abs{m} - M).
\end{equation}

For large disorder, $p \approx 1/2$, random systems can also develop spin glass order at low temperature, which corresponds to an instability in the variance of the magnetization distribution, which is also called the Edwards-Anderson (EA) order parameter 
\begin{equation}
    q_{\rm EA} = \left[m^2\right],
\end{equation}
where the magnetization vanishes ($[m]=0$).
At intermediate values of disorder, there is in principle also the possibility of a magnetized spin glass phase \cite{thouless1986, Carlson1990}, where the magnetization distribution has both finite width ($q_{\rm EA} \neq 0$) and mean ($[m] \neq 0$). 

The schematic phase diagram of the RBIM and its dual on the hyperbolic plane is shown in \autoref{fig:phases_sktech}. Note that for the dual model, we only indicate the phase boundary of the ferromagnetic phase. There could exist a spin-glass phase in principle, but the investigation of that is beyond the scope of this work. 
We also indicate the so called \emph{Nishimori line} \cite{Nishimori1981}, which is defined by the condition 
\begin{equation}
    \exp(2\beta J) = \frac{p}{1-p},
    \label{eq:Nishimori}
\end{equation}
that is the (relative) probability of frustrating a bond due to thermal fluctuations is equal to that of flipping its sign due to the quenched disorder. 
It is known that the multiciritical point in the RBIM lies on the Nishimori line and that the phase boundary of any magnetized phase must be reentrant or vertical, that is no magnetized phase can exist for $p_{\rm N} < p$ \cite{Nishimori1981}.

As indicated, we expect the ferromagnetic phase of the RBIM to have a larger extent than that of its dual, since the additional cocycle degrees of freedom $\eta_{\ell}$ have a finite contribution to the entropy, which is then strictly greater than that of the RBIM.

\section{Methods\label{sec:methods}}

\subsection{High-Temperature Series Expansion\label{sec:series_expansion}}

Our primary means to map out the phase diagram of the random-bond Ising model in hyperbolic space will be to perform high-temperature series expansions of both the susceptibility 
\begin{equation}
    \chi = \beta \frac{1}{N} \sum_{i,j} \left[\expval{\sigma_i \sigma_j} - \expval{\sigma_i}\expval{\sigma_j}\right],
\end{equation}
as well as of the Edwards-Anderson (EA) susceptibility
\begin{equation}
    \chi_{\rm EA} = \beta \frac{1}{N^2} \sum_{i,j} \left[\expval{\sigma_i \sigma_j}^2 - \expval{\sigma_i}^2\expval{\sigma_j}^2\right].
\end{equation}
Coming from a high-temperature, if there is a transition to low-temperatures ferromagnetic phase, the susceptibility $\chi$ at the transition should diverge as a power law
\begin{equation}
    \chi \sim \frac{1}{(T-T_c)^\gamma}
\end{equation}
while the Edwards-Anderson susceptiblity $\chi_{\rm EA}$ can have either a weak singularity or diverge as well \cite{binder1986}. In contrast, if there is a transition into a low-temperature spin-glass phase, the susceptibility $\chi$ will exhibit only a weak singularity (a cusp), while the Edwards-Anderson susceptibility diverges as a power law
\begin{equation}
    \chi_{\rm EA} \sim \frac{1}{(T-T_c)^{\gamma'}}.
\end{equation}

\subsubsection{Biconnected graph expansion of inverse susceptibilities}

\begin{figure}
    \includegraphics[width=0.9\columnwidth]{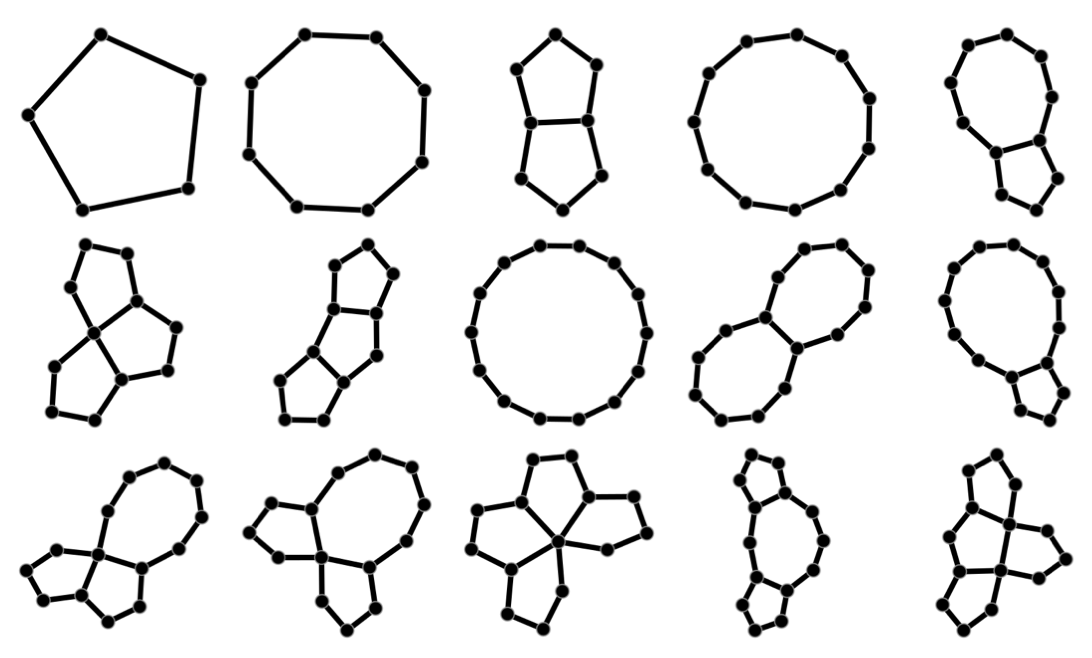}
	\caption{Some small biconnected subgraphs of the $\{5,5\}$-tiling.  Removing a vertex and all its incident edges will leave the graphs connected. Only biconnected graphs contribute to the series expansion.}
	\label{fig:55bicon}
\end{figure}

It turns out that for susceptibilties of the form 
\begin{equation}
    \chi_{k, l} = \beta \frac{1}{N} \sum_{i,j} \left[\expval{\sigma_i \sigma_j}^k - \expval{\sigma_i}^k\expval{\sigma_j}^k\right]^l,
\end{equation}
it is favourable to perform the high-temperature expansion in the {\em inverse} susceptibility. The reason for this is that it can be shown~\cite{singh87} that the only non-trivial contributions come from \emph{biconnected} graphs, that is graphs which stay connected if any of their vertices (and the edges attached to it) are being removed. 
We show the first few graphs that contribute to the susceptibility $\chi=\chi_{1, 1}$ and EA-susceptibility $\chi_{\rm EA}=\chi_{2, 1}$ on the $\{5, 5\}$ lattice in \autoref{fig:55bicon}.

The inverse susceptibility can be expanded in terms of these graphs as a function of both inverse temperature $v=\tanh(\beta J)$ and disorder strength $\mu = 1-2p$. In practice, the variables in the systematic biconnected graph expansion are $w=v^2$ and $\alpha = \mu/v$:
\begin{align}\label{eqn:sus_series}
\tilde{\chi}^{-1}(w, \alpha) = 1\, + \, \sum_{g}\, c(g)\, W(g)
\end{align}
where the sum is over all graphs, $c(g)$ is the coefficient of~$N$ of the number of embeddings of the graph~$g$ into the lattice and~$W(g)$ for each graph is a function of both~$w$ and~$\alpha$. 
Expanding $W$ as a function of inverse temperature $w$, one can show that for each order $n$, the coefficient of $w^n$ is a polynomial in $\alpha$ of order $n$ with integer coefficients. 
For example, the inverse susceptibility on the $\{5, 5\}$ lattice is given by
\begin{align}
\chi^{-1}(w, \alpha) = 1 &- 5 \alpha w + 5 \alpha^2 w^2 - 5 \alpha^3 w^3 + 5 \alpha^4 w^4 \nonumber\\
&+ (10 \alpha + 10 \alpha^2 +10 \alpha^3 + 10 \alpha^4 + 5 \alpha^5) w^5 \nonumber\\
&+ \order{w^6}.
\end{align}
Note that for $\alpha = 1$ (that is $v=\mu$), we obtain the series on the Nishimori line up to order $\order{w^n}=\order{v^{2n}}$.

For more details and a derivation of \autoref{eqn:sus_series} see Ref.~\onlinecite{singh87}.

\subsubsection{Analysis of the series}

We analyze the generates series $\tilde{\chi}(w, \alpha)$, usually for fixed $\alpha$ as a function of $w$, using \emph{first-order homogeneous integrated differential approximants (FO-IDAs)}.
One reason to choose FO-IDAs over simpler methods is that they are known to be less biased towards the lower-order coefficients of the expansion~\cite{singh2}.
This is important, as the most relevant contributions on a $\{r, s\}$ tiling come from graphs with at least~$r$ edges.

The analysis using FO-IDAs proceeds as follows:
For fixed disorder strength $\alpha$, we assume that the series $\tilde{\chi}$ is the solution of a first-order differential equation of the form
\begin{equation}\label{eqn:IDAdef}
Q_L(w) \frac{d \tilde{\chi}(w)}{d v} + R_M(w)\, \tilde{\chi}(w) + S_T(w) = 0
\end{equation}
where $Q_L$, $R_M$ and $S_T$ are polynomials of degree $L$, $M$, $T$, respectively.
By equating the series order-by-order with the coefficients of \autoref{eqn:IDAdef} we obtain a linear system of equations in the coefficients of the polynomials $Q_L$, $R_M$ and $S_T$.
It can be shown that for any root $w_c$ of the polynomial~$Q_L$, a solution of \autoref{eqn:IDAdef} has an algebraic singularity of the form $(w-w_c)^{-\gamma}$ \cite{oitmaa2006}.
The exponent of the singularity is given by
\begin{align}\label{eqn:crit_exp}
\gamma = \frac{R_{M}(w_c)}{Q'_L(w_c)} .
\end{align}
Generally, the results for $w_c$ and $\gamma$ will depend on the choice of degrees $L$, $M$ and $T$.
If we have the series up to order $N$ then we can choose all possible values satisfying $L+M+T \leq N-2$.
Following~\cite{singh2} we exclude approximants if one of the following is true
\begin{itemize}
	\item a root of $R_M$ is close to $w_c$, giving rise to a small estimate of $\gamma$
	\item a complex root of $Q_L$ with small absolute value smaller than $w_c$ is close to the real axis
\end{itemize}

We observe that the convergence of the series is very good, since the approximants for different choices of $L$, $M$ and $T$ are all close.

\subsection{Monte Carlo Simulations}

To corroborate our results from the series expansion and to compute additional observables, we also perform classical Monte-Carlo simulations for some sets of parameters. 
To compute the disorder average $\left[\dots\right]$, we perform Monte-Carlo simulations for using 1000 disorder realizations $\left\{J_{ij}\right\}$. For each realization, we simulate two independent copies $\{ \sigma_j^{(1)} \}$, $\{ \sigma_j^{(2)} \}$ of the system.

\subsubsection{Equilibration in the (possible) presence of glassiness}

Since it is know that there is no spin glass behavior on the Nishimori line \cite{Nishimori1981}, we expect that a standard local Metropolis-Hastings algorithm is sufficent to equilibrate the system at temperatures $T > 2J\log[p/(1-p)]^{-1}$. 
When approaching the spin glass phase, the local algorithm suffers from a drastic slowdown. Nevertheless, we are able to study the spin glass transition since for that we do not need to equilibrate the system deep inside the glassy phase. To make sure that the system is actually equilibrated, we keep track of the autocorrelation time of all relevant observables (computed via binning analysis~\cite{ALPSCore}) to ensure that we equilbrate the system for at least $10$ times as long as the largest equilibration time in the system and that we take $5000$ independent samples per temperature value for each observable

\subsubsection{Finite size scaling}
Due to the absence of a unique linear dimension in the compactifications of the hyperbolic plane, we perform finite size scaling as a function of the number of sites~$N$. This was initially proposed for a fully connected model~\cite{Botet1982} and has been used for hyperbolic lattices with open boundary conditions \cite{Shima2006} as well as in our study of the pure Ising model in the hyperbolic plane \cite{breuckmann2020}.
The main idea is that a quantity $A$, close to criticality, follows a scaling form
\begin{equation}
    A \sim \abs{T-T_c}^a \, F\left(N/N_c\right)
\end{equation}
with a correlation number $N_c$.
Assuming that a corresponding system of finite dimension $d = d_c$, where $d_c$ is the upper critical dimension, has the same scaling behavior as its hyperbolic sibling, it follows that 
\begin{equation}
    N_c = \sim \abs{T-T_c}^{-\mu},
\end{equation}
with the critical exponent 
\begin{equation}
   \mu = \nu_{\rm MF} \, d_c,
\end{equation}
and $\nu_{\rm MF}$ is the mean-field value of the critical exponent of the correlation length $\xi$.

\subsubsection{Observables}

To map out the phase diagram and compute critical exponents, we study a number of observables, all of which are related to either the magnetization
\begin{align}
    m^{(\alpha)} = \frac{1}{N} \sum_{j} \sigma_j^{(\alpha)} \label{eq:mag}
\end{align}
or the Edwards-Anderson order parameter
\begin{equation}
    q = \frac{1}{N} \sum_{j} \sigma_j^{(1)} \sigma_j^{(2)}.
\end{equation}
First, to determine the location of the critical point and the critical exponent of the correlation number $\mu$, we compute the binder cumulants
\begin{align}
    g &= 1 - \frac{\left[\expval{m^4}\right]}{\left[\expval{m^2}^2\right]}, \label{eq:binder}\\
    g_{\rm EA} &= 1 - \frac{\left[\expval{q^4}\right]}{\left[\expval{q^2}^2\right]} \label{eq:binderEA}
\end{align}
which, for different system sizes, cross at the transition to a magnetized and a spin glass phase respectively. The best estimate for the transition temperature $T_c$ and the exponent $\mu$ is given by performing a data collapse, using the fact that close to the transition the respective cumulant is given by
\begin{equation}
    g = G\left(N^{1/\mu}(T-T_c)\right),
\end{equation}
with some universal scaling function $G$.

For both order parameters, we also compute the corresponding susceptibilties 
\begin{align}
   \chi &= \beta N\left( \left[\expval{m^2}\right] - \left[\expval{m}\right] \right), \label{eq:sus}\\
   \chi_{\rm EA} &= \beta N\left( \left[\expval{q^2}\right] - \left[\expval{q}\right] \right). \label{eq:susEA}
\end{align}
Again, the best estimate for $T_c$, $\gamma$ and $\mu$ are obtained by performing a data collapse, since close to the transition the susceptibility is given by
\begin{equation}
   \chi =  N^{\gamma/\mu} S\left(N^{1/\mu}(T-T_c)\right),
\end{equation}
with some universal scaling function $S$.

\section{Results for the RBIM\label{sec:rbim}}

\subsection{Phase diagram on the \{5, 5\} lattice}

To study general features of the phase diagram of the  RBIM in hyperbolic space as well as to assess the reliability of the high-temperature series expansion (HTSE) in the presence of disorder, we first map out the phase diagram of the model on the $\{5, 5\}$ lattice in detail, using both HTSE as well as Monte-Carlo simulations.

\begin{figure}
    \centering
    \includegraphics{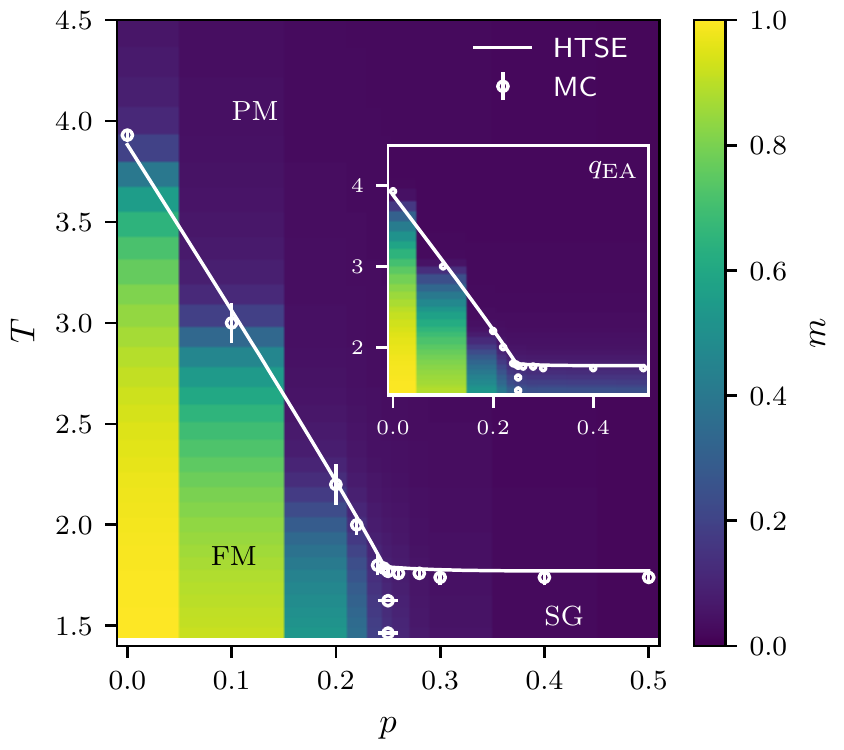}
    \caption{Phase diagram on the $\{5, 5\}$ lattice as a function of temperature $T$ and disorder strength $p$. We show both the magnetization $m$ as well as the Edwards-Anderson order parameter $q_{\rm EA}$ (inset) obtained from Monte-Carlo (MC) simulations of a $N=1920$ system. We superimpose this with the phase boundaries obtained from the high-temperature series expansion (HTSE) and MC (see main text for details).}
    \label{fig:phases55}
\end{figure}

The phase diagram of $\{5, 5\}$ is obtained from HTSE and MC simulations is shown in \autoref{fig:phases55}. Compared to the RBIM on the euclidean square ($\{4, 4\}$) lattice, we find a much larger ferromagnetic phase and a extended spin glass phase. In contrast to the Bethe lattice, here we do not find evidence for a magnetized spin glass phase, although our low-temperature data is not good enough to rule out a very small extent.

Turning to explain our results in more detail, in \autoref{fig:phases55} we show both, the magnetization $m$ as well as the Edwards-Anderson order parameter $q$ (in the inset) as obtained from a MC simulation with system size $N = 1920$. While the magnetization is nonzero only in the ferromagnetic phase, the EA order parameter is nonzero in both the ferromagnet and the spin glass. 
We superimpose these plots with the critical points obtained using finite-size scaling of the MC data (open circles) and with the critical lines obtained from HTSE of the (EA-) susceptibility (solid lines). In both methods, we can distinguish the transition from the paramagnet to a ferromagnetic phase and that to a spin glass phase reliably. In the finite size analysis of the MC data, a transition to the ferromagnetic phase is signaled by a crossing of both the binder cumulant of the magnetization, $g$ [\autoref{eq:binder}] as well as a crossing of the binder cumulant of the Edwards-Anderson order parameter, $g_{\rm EA}$ [\autoref{eq:binderEA}]. In contrast, at the transition to a spin glass phase, only $g_{\rm EA}$ shows a crossing while $g$ does not, since the magnetization $m$ vanishes in the spin glass. Finite size scaling along the Nishimori line indicates a transition at $p_{\rm N} = 0.247 \pm 0.02$ and finite size analysis as a function of temperature at constant disorder shown a transition into a ferromagnet for $p \lessapprox p_{\rm N}$ and a transition into a spin glass for $p \gtrapprox p_{\rm N}$, making the Nishimori point the multicritical point. 

This result is corroborated by HTSE analysis. Here, a transition to the ferromagnet (spin glass) is signaled by the divergence ferromagnetic (EA-) susceptibility $\chi_{(\rm EA)}$. Note that since the non-divergent susceptibility at both transitions typically also has a weak singularity (a cusp), series analysis normally predicts a divergence for both susceptibilities, but at different critical temperatures. In practice, we distinguish the two transitions by the fact which susceptibility is predicted to diverge at a larger temperature. Along the Nishimori line, that is $\alpha = 1$ in \autoref{eqn:sus_series}, the two susceptibilities are equal and HTSE yields a critical point $w_c = 0.256456 \pm \num{8.6e-6}$, which corresponds to $p_{\rm N} = 0.246793 \pm \num{4.2e-6}$. For $\alpha < 1$ we find a transition to a ferromagnetic phase while for $\alpha > 1$ we find a transition into a spin glass phase, again suggesting that the Nishimori point is indeed the multicritical point of the model.

\subsection{Phase boundaries for different tilings: coordination vs curvature}

\begin{figure}
    \centering
    \includegraphics[width=\columnwidth]{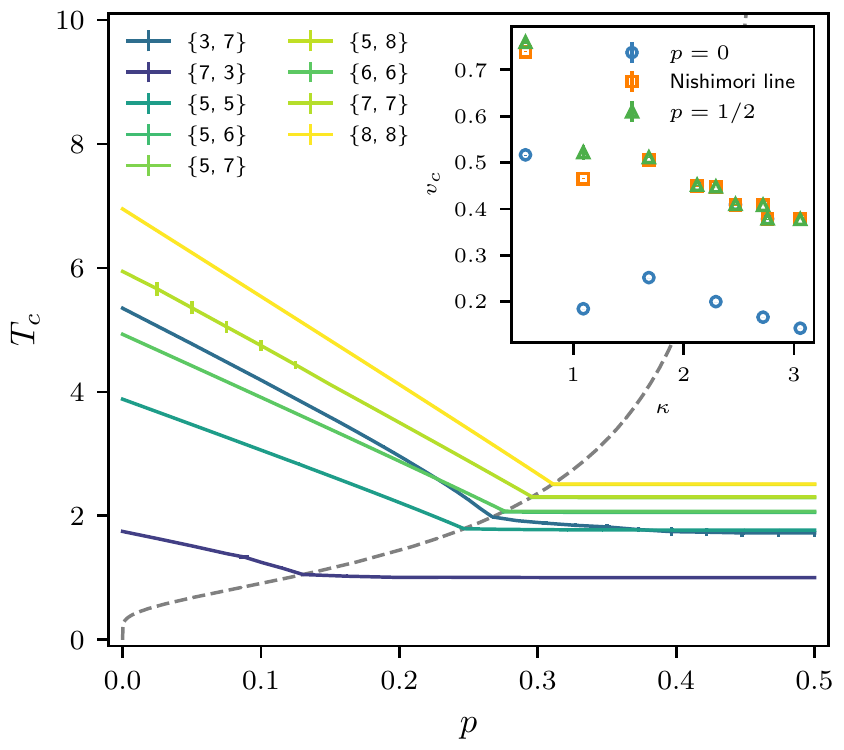}
    \caption{Critical temperature $T_c$ obtained from high-temperature expansion, for different tilings of the hyperbolic plane. The inset shows $v_c = \tanh(J/T_c)$ as a function of curvature $\kappa$ for the pure model ($p=0$), along the Nishimori line ($(1-p)/p = e^{-2\beta J}$) and for the spin glass boundary ($p=1/2$).}
    \label{fig:phases_curv}
\end{figure}

We now use the high-temperature expansion to study how the paramagnet-ferromagnet and paramagnet-spin-glass phase boundaries vary for different tilings $\{r, s\}$. For low disorder, the critical temperature is mostly controlled by the coordination number $s$ and for $p=0$ even agrees quantitatively with that of the Bethe lattice with the same coordination \cite{breuckmann2020}.
Qualitatively, this behaviour can be understood by considering that the transition into the ferromagnet at low disorder is driven by a competition between  and the entropy of the paramagnet and internal energy of the ferromagnetic state
\begin{equation}
    E_{\rm FM} = \frac{sN}{2}\left[J_{ij}\right],
\end{equation}
which is proportional to the coordination number $s$. This means that with larger~$s$, the ferromagnet becomes more favorable at larger temperatures. 
As disorder is increased however, $[J_{ij}]$ also increases (approaching zero from a negative value) and so does the importance of~$s$ as a control parameter for the transition temperature. 
Finally, $[J_{ij}] \to 0$ as $p\to \frac{1}{2}$ and the critical temperature becomes a monotonic function of the curvature $\kappa$, as seen in the inset of  \autoref{fig:phases_curv}.

\subsection{Critical Behaviour}

\begin{table}
    \centering
    \begin{tabular}{c|c|c|c}
            & $p=0$ & Nishimori Line & $p=1/2$\\\hline
        $\mu$ & 2 & $3.0 \pm 0.1$ & $2.0 \pm 0.1$ \\
        $\gamma$ & $1.000001 \pm 0.000005$ & $1.0003 \pm 0.0008$ & -\\
        $\gamma_{\rm EA}$ &  - & $1.0003 \pm 0.0008$ & $1.0011 \pm 0.0025$ \\
        $\beta$ &  $0.46 \pm 0.05$ & $1.00 \pm 0.05$ & -
    \end{tabular}
    \caption{Critical exponents on the $\{5, 5\}$ lattice along different scaling axes. We estimate the correlation volume exponent, $\mu$, from finite size analysis of the binder parameter $g$. For the susceptibility exponents $\gamma$ and $\gamma_{\rm EA}$ the best estimates are obtained via HTSE analysis.  }
    \label{tab:exponents}
\end{table}

\begin{figure}
    \centering
    \includegraphics{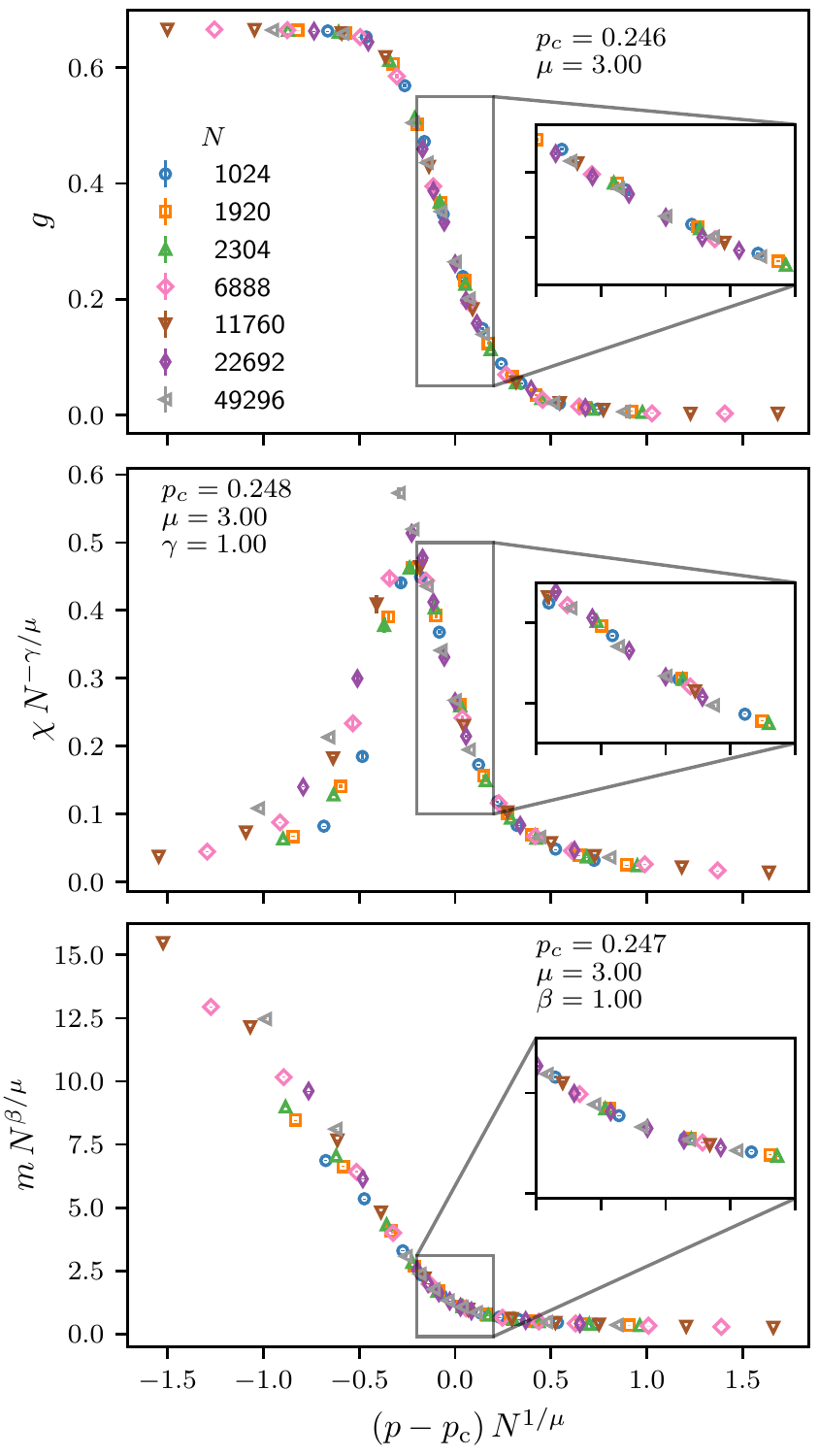}
    \caption{Finite size scaling collapse of the Binder cumulant~$g$ [\autoref{eq:binder}], the susceptibility [\autoref{eq:sus}], and the magnetization $m$ [\autoref{eq:mag}] of the random bond Ising model on the $\{5, 5\}$ lattice along the Nishimori line [\autoref{eq:Nishimori}].}
    \label{fig:collapse-N}
\end{figure}

In \autoref{tab:exponents}, we show out best results for the critical exponents for the $\{5, 5\}$ lattice for different scaling axis (with the $p=0$ results taken from Ref. \onlinecite{breuckmann2020}).
The best results are typically obtained from the HTSE, except for the exponent $\mu$ of the correlation volume, which we compute by finite size analysis of the Monte-Carlo data. 
In all cases were results from both methods are available, they are compatible within errors. 
The best finite-size scaling collapse of the Monte Carlo data along the Nishimori line is shown in \autoref{fig:collapse-N}.
The best collapse is obtained for slightly different values of $p_c$ for the susceptibility and the binder cumulant, which we attribute to finite size effects.

The results in \autoref{tab:exponents} are all compatible with the mean-field expectation, except for the exponents $\mu = 3$ and $\beta = 1$, observed along the Nishimori line. This is because as established in \autoref{sec:rbim}, the Nishimori line passes through the multicritical point, which generally shows distinct critical behavior even in (effectively) infinite dimensions. 
Note that still, the exponents are consistent with the hyperscaling relation
\begin{equation}
    \mu = 2\beta + \gamma
\end{equation}
Note that the specific heat does not develop a power-law singularity for any of the transitions considered and hence we do not present a critical exponent $\alpha$.

\section{Results for the dual-RBIM\label{sec:dual-rbim}}

In this section, we present results of Monte-Carlo simulations of the dual random-bond Ising model (dual-RBIM). We present evidence that this model exhibits a strongly first-order transition as a results of its cocycle degrees of freedom and numerically verify that for $p=0$, the critical temperature of this transition is indeed the Kramers--Wannier dual to the critical temperature of the Ising model on the dual lattice.

\subsection{Dual Ising model}

\begin{figure}
    \centering
    \includegraphics{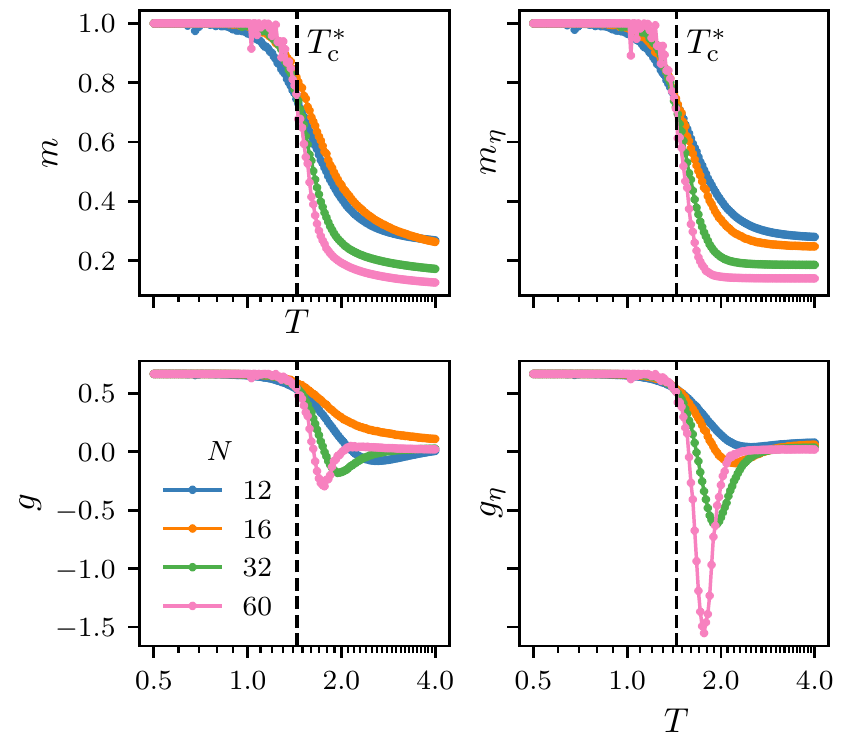}
    \caption{Evidence for a strongly first-order phase transition of the pure dual Ising model (that is \autoref{eq:H-dual-ising} with $p=0$) on the $\{5, 5\}$ lattice. We show the vertex magnetization $m = \expval{\sigma_j}$, its Binder cumulant $g$ as well as the loop magnetization $m_{\eta} = \expval{\eta_\ell}$ and its Binder cumulant $g_{\eta}$ as a function of temperature.}
    \label{fig:dual-mc}
\end{figure}

In \autoref{fig:dual-mc}, we show results from Monte-Carlo siumulations of the dual Ising model, that is \autoref{eq:H-dual-ising} with $p=0$, on the $\{5, 5\}$ lattice. We show the average vertex magnetization $m = \expval{\sigma_j}$ and \emph{loop} magnetization $m_{\eta} = \expval{\eta_\ell}$, as well as the Binder cumulants $g$ and $g_\eta$ for vertex and loop magnetization respectively. The fact that the magnetizations for different system sizes cross at a single point, together with the pronounced dip of the Binder cumulants just before the transition are evidence that both quantities undergo a strongly first-order transition. This result is rather surprising, given that the magnetization of the Ising model on the same lattice undergoes an ordinary second-order transition, and the two models are related by the Kramers-Wannier duality.

Since the $\{5, 5\}$ lattice is self dual and the Kramers--Wannier duality [\autoref{eq:krammers-wannier-T}] is exact at $p=0$, we expect the transition to occur at a critical temperature dual to the the critical point of the Ising model. Subsituting $T_c = 3.93$ \cite{breuckmann2020} into \autoref{eq:krammers-wannier-T} yields $T_{\rm c}^* \approx 1.44$, which we indicate in \autoref{fig:dual-mc} by a vertical dashed line and is in good agreement with the position of the crossing of both Binder cumulants and magnetizations.

\begin{figure}
    \centering
    \includegraphics{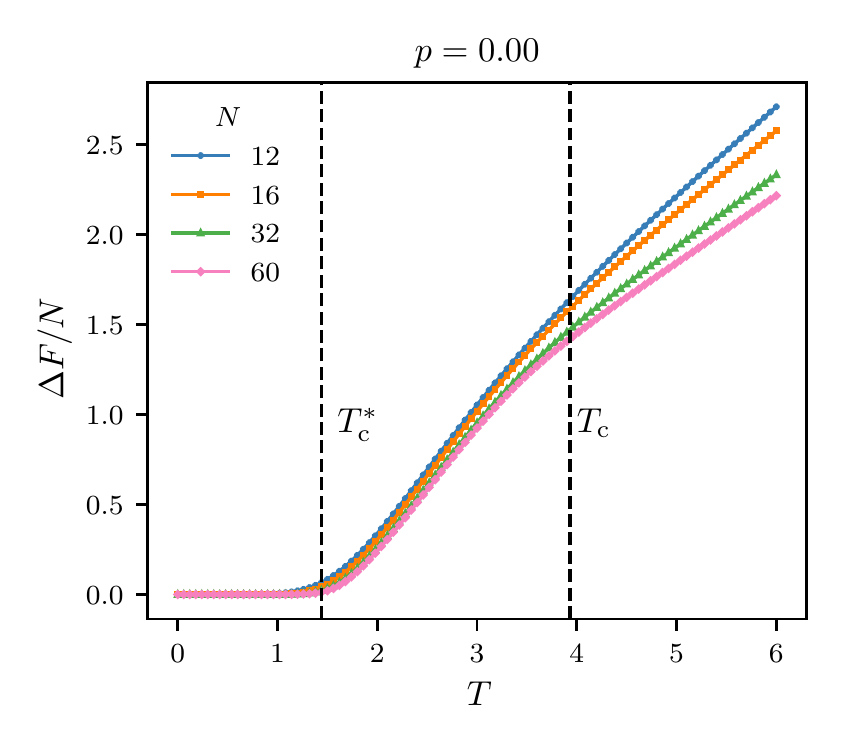}
    \caption{Free energy difference [\autoref{eq:fdiff}] between the Ising model and its dual on the $\{5, 5\}$ lattice.}
    \label{fig:dual-wl-pure}
\end{figure}

To corraborate the above findings, we also implement the Wang-Landau algorithm \cite{wang_landau2001, wang_landau2001b, schulz2003, belardinelli2007} and compute the free energy difference of the Ising model and its dual that is
\begin{equation}
    \label{eq:fdiff}
    \Delta F(T) = \log[Z_{\rm tot}(T)] - \log[Z_{0}(T)].
\end{equation}
Here, $Z_{\rm tot}$ is the partition function of the dual Ising model, that is it includes a sum over all cocycle variables~$\eta_\ell$ (therefore the subscript `tot'). $Z_0$ is the partition function of the Ising model on the same lattice, that is we fix $\eta_\ell = 1$ for all $\ell$.
Because of the latter relation between $Z_{\rm tot}$ and $Z_0$, we have $\Delta F > 0$ for all~$T$. In the ordered phase of the dual Ising model, the difference vanishes since the sum over cocycle variables does not contribute. This is shown in \autoref{fig:dual-wl-pure}. The quantity $\Delta F$ also has the advantage of indicating both phase transitions in one observable, since the free energy of the Ising model shows a visible kink at $T_{\rm c}$. Both critical temperatures are again indicated in the figure by vertical dashed lines.

\subsection{Dual random bond Ising model} 

\begin{figure}
    \centering
    \includegraphics{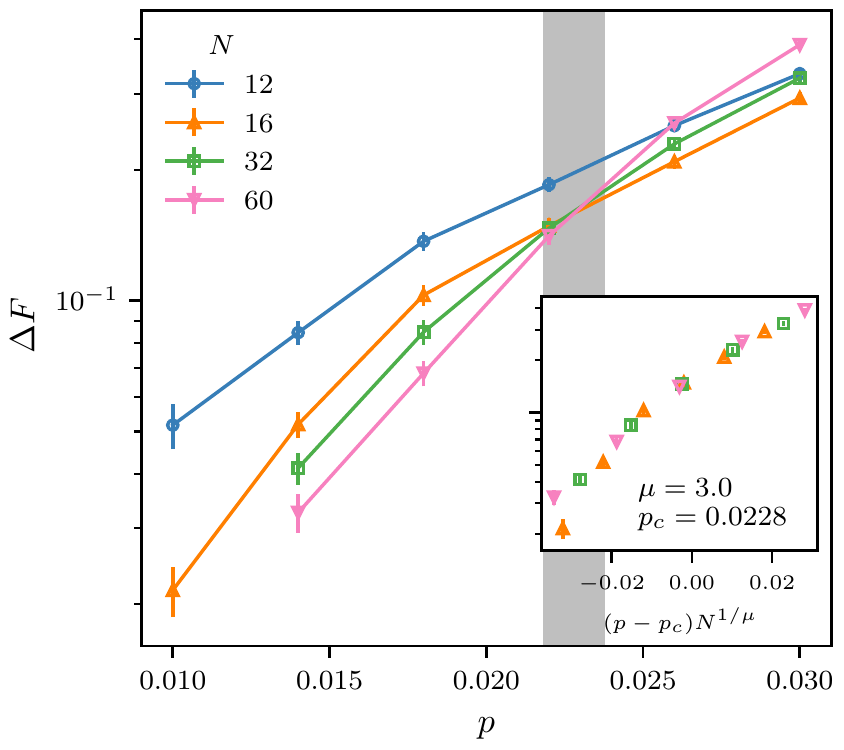}
    \caption{Free energy difference [\autoref{eq:fdiff}] between the random bond Ising model (RBIM) and the dual-RBIM on the $\{5, 5\}$ lattice along the Nishimori line. The shaded region indicates the location of the Nishimori point $p_{\rm N} = 0.0228 \pm 0.0010$. The inset shows the best data collapse, assuming the same correlation exponent $\mu$ as in the RBIM.}
    \label{fig:dual-wl-Nishimori}
\end{figure}

In the case of the random model, the dual-RBIM is not exactly dual to the RBIM and hence we have a a-priori guess for the location of the critical point. Additionally, as is already the case in the pure model, the strongly first order nature of the transition complicates its numerical investigation. We find that single-spin flip Monte-Carlo is unreliable even for small system sizes. However the Wang-Landau algorithm is still converging and hence we can infer the location of the critical point from the free energy difference [\autoref{eq:fdiff}]. The difference for $T < T_c^*$ vanishes as a function of system size and diverges as a function of system size for $T_c^* < T$ respectively. 
In \autoref{fig:dual-wl-Nishimori}, we show $\Delta F$ as a function of disorder strength~$p$ along the Nishimori line [\autoref{eq:Nishimori}]. The data is consistent with a transition at $p_{\rm N}^* = 0.0228 \pm 0.001$, which is indicated in the figure by a shaded area. The inset shows the best data collapse assuming the same correlation exponent $\mu = 3$ as in the RBIM.

Substituting the value of $p_{\rm N} = 0.246793 \pm \num{4.2e-6}$ obtained from the high-temperature expansion of the RBIM (see \autoref{sec:rbim} for details) into the duality relation conjectured by Nishimori (\autoref{eq:duality-conjecture}) and solving for $p_{\rm N}^*$ yields a value of  $p_{\rm N}^* = 0.029891 \pm \num{2e-6}$. As observed for the RBIM on a range of euclidean lattice geometries \cite{takeda2005exact} this is somewhat close to our numerical result but not compatible within error bars.

\section{Quantum Error Correction\label{sec:qec}}

Quantum error correcting codes are used in quantum computation to reduce the effects of decoherence.
Certain infinite families of codes, together with associated quantum error correction protocols, can be shown to have a \emph{threshold}.
A threshold is a critical value of a noise parameter, below which the error correction protocol succeeds with probability approaching 1 with increasing code sizes.

It was argued in \cite{dennis2002topological,wang2003confinement} that the threshold of the toric code corresponds to the phase transition point along the Nishimori line of the RBIM on the square-grid $\{4,4\}$.
In~\cite{kubica2018three} it was proved that this is indeed the case for quantum codes which encode a finite number of qubits.
In~\cite[Section~IV-C]{chubb2021statistical} it was mentioned that the statistical mechanical models associated to quantum codes which encode an extensive number of qubits may exhibit multiple phase transitions.
This behaviour was studied in \cite{kovalev2018numerical}.

The quantum codes associated to the hyperbolic RBIM are called \emph{hyperbolic surface codes}~\cite{breuckmann2016constructions,breuckmann2017hyperbolic,conrad2018small}.
These codes do encode an extensive number of qubits, so that the proofs of \cite{kubica2018three,chubb2021statistical} do not apply to them.
In \cite{jiang2019duality} the authors consider the hyperbolic RBIM and give a condition sufficient for error correction to be possible, which is equivalent to $\Delta F \to 0$, where $\Delta F$ is the free energy difference of the RBIM and the dual-RBIM, see \autoref{eq:fdiff}. 
Hence, assuming all logical operators are equivalent, the phase transition of what we call the ``dual-RBIM'' along the Nishimori line corresponds exactly to the maximum likelihood decoding threshold of the hyperbolic surface code under independent bit- and phase-flip noise
\begin{equation}
    p_{\rm th, ML} = p_{\rm N}^* = 0.0228 \pm 0.0010.
\end{equation}
This can be compared to the threshold when using a minimum-weight perfect-matching decoder, which is $p_{\rm th, MWPM} \approx 0.0175$ \cite{breuckmann2017homological}.
Using an optimal decoder rather than MWPM hence increases the threshold by about 27\%.

\section{Conclusion\label{sec:conclusion}}

To summarize, we have presented an in-depth study of the random bond Ising model (RBIM) on the hyperbolic plane as well as the model that is its Kramers--Wannier dual in the absence of disorder. Resolving a conundrum raised in earlier work \cite{Rietman1992, breuckmann2020}, we showed that this ``dual-RBIM'' is different from the RBIM even on self-dual lattices due to the extensive number of nontrivial cocycles of hyperbolic lattices. Combining high-temperature expansion techniques and Monte-Carlo techniques, we mapped out the phase diagrams of both models, establishing the existence of a spin-glass phase with the Nshimori point as the tricritical point. Studying the critical properties of the high-temperature transitions, we showed that with the exception of the multicritical point, all transitions are mean-field in nature.
We verified the duality of both models explicitly in the disorder-free case and showed that the extended duality as conjectured by Takeda, Sesamoto and Nishimori \cite{takeda2005exact} is fulfilled only approximately.
Finally, we commented on the relation of the above findings to the decoding of hyperbolic surface codes and argued that the critical disorder along the Nishimori of what we call the dual-RBIM corresponds to the maximum-likelihood decoding threshold of hyperbolic surface codes under independent bit- and phase-flip noise. This generalizes the statistical mechanics mappings of the decoding of zero-rate quantum codes \cite{dennis2002topological, chubb2021statistical, kubica2018three} to quantum codes with \emph{finite} rate.

This work open up multiple interesting ares for future work. For example, beyond the scope of the current paper was a detailed investigation of the nature of the spin-glass phase in hyperbolic space and in particular its fate in the dual-RBIM. 
Moreover, a detailed investigation of the phase space structure of the dual model could yield valuable insights into the decoding of finite-rate quantum codes.

\subsection*{Acknowledgements}
We thank Ananda Roy for many helpful discussions in the early stages of this project. We also thank Leonid Pryadko for many helpfull comments and suggestions on this work. 
We thank Aleksander Kubica, Sounak Biswas, Rajiv Singh and Roderich Moessner for fruitful discussions, and also Philippe Suchsland, Dmitry L. Kovrizhin and Peng Rao for helpful comments on the manuscript.
BP acknowledges support by the Deutsche Forschungsgemeinschaft  under grants SFB 1143 (project-id 247310070) and the cluster of excellence ct.qmat (EXC 2147, project-id 390858490). 
NPB acknowledges support through the EPSRC Prosperity Partnership in Quantum Software for Simulation and Modelling (EP/S005021/1).

\appendix
\section{Kramers--Wannier Duality as Fourier Transformation}
The Kramers--Wannier duality is in fact a Fourier transformation of the partition function.
In this section we provide the formal argument.

First, let us rewrite the partition function in terms of homological algebra.
To this end, we require some definitions.
Let $$C_0 = \left\lbrace \sum_{v\in V} a_v v \mid a_v \in \mathbb{Z}_2 \right\rbrace$$ be the vector space containing formal linear combinations of vertices with coefficients in $\mathbb{Z}_2$ and similarly $$C_1= \left\lbrace \sum_{e\in E} a_e e \mid a_e \in \mathbb{Z}_2 \right\rbrace$$ the $\mathbb{Z}_2$-vector space spanned by the edges.
The coboundary operator $\delta_0$ is represented by a $\mathbb{Z}_2$-matrix whose rows are labeled by edges and columns labeled by vertices and $(\delta_0)_{e,v}=1$ if $v\in e$ and $0$ otherwise.
We may think of the coboundary operator $\delta_0$ as a discrete version of the gradient operating on $\mathbb{Z}_2$-scalar fields $\phi \in C_0$.
The boundary operator is defined as $\partial_1 = \delta_0^{tr}$.
As in \autoref{eq:low-T-z}, we observe that the sum $\sum_{i\sim j} \sigma_i \sigma_j$ can be rewritten as $|E|-2|\delta_0 \phi|$, where $|\cdot |$ is the Hamming weight.
To simplify notation, we introduce the function $f(c) = \exp(KN-2|c|)$.
Hence, we can express the partition function as a sum over all gradients of $\mathbb{Z}_2$-fields.
In order to Fourier transform $f$, we observe that the characters of $C_i$, interpreted as abelian groups, are given by $\chi_d(c) = (-1)^{\langle c,d \rangle}$.
Hence, we obtain:
\begin{subequations}
\label{eq:fourier}
\begin{align}
\sum_{\phi \in C_0}  f(\delta_0 \phi)  &= \sum_{\phi \in C_0} \frac{1}{\sqrt{|C_1|}} \sum_{\xi \in C_1} (-1)^{\langle \delta_0 \phi, \xi \rangle} \hat{f}(\xi)\\
&= \frac{1}{\sqrt{|C_1|}} \sum_{\xi \in C_1} \hat{f}(\xi) \sum_{\phi \in C_0} (-1)^{\langle \phi, \partial_1 \xi \rangle} \\
&= \frac{|C_0|}{\sqrt{|C_1|}} \sum_{\gamma \in Z_1} \hat{f}(\gamma)\\
&= 2^{N-|E|/2} \sum_{\gamma \in Z_1} \hat{f}(\gamma)
\end{align}
\end{subequations}
In the second equation we used $\delta_0^{tr} = \partial_1$ and in the third equation we used that the sum over character values is zero, unless it is the trivial character.
Note that Fourier transforming turned the partition function from a sum over coboundaries $\delta_0 \phi \in B^1$ into a sum over cycles $\gamma \in \operatorname{ker} \partial_1$.

The Fourier transformed function $\hat{f}$ can be expressed as follows.
\begin{subequations}
\label{eq:ftransformed}
\begin{align}
\hat{f}(\gamma) &= \frac{1}{\sqrt{|C_1|}} \sum_{c\in C_1} f(c) (-1)^{\langle \gamma, c\rangle}\\
&= \frac{1}{2^{|E|/2}} \prod_{e \in E} \left( \exp(K) + (-1)^{\gamma_e} \exp(-K)\right)\\
&= \frac{\left(\cosh K \right)^{|E|}}{2^{|E|/2}}  \prod_{e \in E} \left[ 1 + (-1)^{\gamma_e} + (1-(-1)^{\gamma_e}) \tanh(K) \right] \\
&= 2^{|E|/2} (\cosh K)^{|E|}  \left(\tanh K\right)^{|\gamma|}
\end{align}
\end{subequations}
Substituting \autoref{eq:ftransformed} in \autoref{eq:fourier} gives \autoref{eq:kw-high-T}.

The argument given here can be written more abstractly as a Pontrjagin duality between the chain complexes associated with the lattice and its dual, see \cite[Section 4]{freed2018topological}.
We also note that the Kramers--Wannier duality can be seen as a special case of more general dualities derived in algebraic geometry \cite{ikeda2018topological}

\bibliography{references.bib}

\end{document}